\documentclass[a4paper,12pt]{article}

\usepackage[latin2]{inputenc}
\usepackage{graphicx}
\usepackage{amstext}
\usepackage[]{ntheorem}
\usepackage{fancyhdr}
\usepackage{amsfonts,dsfont}
\usepackage{amsmath,amssymb}
\usepackage{multirow}
\usepackage{t1enc}
\usepackage{authblk}
\usepackage{color}
\usepackage{algorithm}
\usepackage{algpseudocode}
\usepackage{subcaption}
\usepackage{hyperref}
\title{Solidarity in natural gas storage: A  potential allocation mechanism of stored quantities among several players during times of crisis}

\author[1,2]{D\'{a}vid Csercsik}

\author[3]{Anne Neumann}

\affil[1]{Centre for Economic and Regional Studies, T\'oth K\'alm\'an u. 4., H-1097 Budapest, Tel.: +36-1 309 26 52, Fax: +36-1 319 31 36 Email: \texttt{\small csercsik.david@krtk.hu}}

\affil[2]{P\'{a}zm\'{a}ny P\'{e}ter Catholic University, Faculty of Information Technology, P.O. Box 278, H-1444 Budapest}

\affil[3]{Department of Industrial Economics and Technology Management, Norwegian University of Science and Technology, Trondheim, Norway.}

\begin{document}

\maketitle

\abstract{The recently experienced disruptions in the EU's energy supply pointed out that supply crises pose a real thread and the member states must be better prepared do deal with the related challenges.
According to the current practice, member states fill their gas storages independently, while it is not clear how solidarity could be put into practice in the future, i.e. how the accumulated reserves of one or more members may be potentially redistributed to help others in need. In this paper we propose some possible guidelines for a potential solidarity framework, and formalize a game-theoretic model in order to capture the basic features of the problem, considering the related uncertainty of the future conditions related to gas storage levels and possible transmission bottlenecks as well. The proposed mechanism of supply-security related cooperation is based on voluntary participation, and may contribute to the more efficient utilization of storage capacities. Via the computational model we demonstrate the operation of the proposed framework on a simple example and show that under the assumption of risk-averse participants, the concept exhibits potential.}

\section{Background}
\label{sec_Introduction}

%\subsection{Gas supply security concerns in the EU}
In 2022, the EU experienced its first energy crisis ever since the oil price crisis during the 1970ies. One of the key fuels in European energy consumption is natural gas, of which a significant part is imported. Natural gas can be stored and thus potentially compensate for seasonal demand imbalances. During summer (usually with lower prices than in winter) natural gas is accumulated in storage facilities from where it is withdrawn during the winter period to meet higher demand for heating.
Storage typically provides 25-30\% of natural gas consumed in the EU during winter, however the extent to which storage is used in EU member states differs significantly. This is due to heterogeneous size of storage infrastructure capacities, the corresponding difference in technically available working gas capacities, and the different amplitudes of seasonal swing (i.e. less need for heating in southern countries).

During the first year of war it seemed possible (if not likely) that most of European gas storages would be depleted during the winter period and several European countries would not be able to re-fill during summer 2023 to levels necessary to ensure security of gas supply for Winter 2023/2024 \cite{ENTSOG2022}.
As of date there exist no official guidelines on cooperation mechanisms between Member States and efficient utilization of underground storage facilities in order to ensure security supplies during crisis. It is worth noting that there is no homogeneous use of either natural gas in the energy mix across, nor an even distribution of capacities across Member States. Whereas storage capacities were filled at almost 95\% on average in the EU, 14\% were left unfilled in Hungary at the beginning of November 2022 \footnote{Note that in June 2021 Hungary's capacities were filled at almost 58\%, the average European level was only 38\% (Gas Infrastructure Europe, AGSI}. Hence, in the case of emergency and the need for solidarity it would be useful to have a transparent mechanism in place. Such mechanism  should be based on optional participation and designed such that it is sufficiently attractive (beneficial) for decision makers to cooperate.
%\subsection{Related literature}
%\label{subsec_related_literature}
The related literature for such a mechanism is scarce. While there are several examples for the application of game theoretic methods for oil and gas related problems (see e.g. \cite{roson2015bargaining,hubert2015pipeline,araujo2018game,jafarzadeh2021possibility,Toufighi2022}), quantitative methods related to the economics of storage-usage are - to the best of our knowledge - lacking.
Although there exist models describing reservoir operation in the context of economic implications of CO$_2$ storage \cite{zhang2007system,schaef2014co2}, the literature on the concept of storage facility sharing is very scarce. \cite{holland2009strategic,holland2013equilibrium} consider the sharing of a single reservoir while taking into account characteristic technological features, but there are no analyses of sharing of multiple storages in a networked setting under uncertainty.
\cite{kiely2016market} investigates accumulation and redistribution whilst taking into account risk-pricing related to natural gas storage. The authors in \cite{janjua2022asymmetric} analyse redistribution and present an asymmetric hybrid bankruptcy and Nash bargaining model for natural gas distribution. \cite{schitka2014applying} and \cite{rey2020regional}
capture the simultaneously competitive and cooperative aspects of gas-related issues.

%\subsection{Motivation and aim}
%\label{subsec_motivation_and_aim}
The goal of this paper is to provide a framework for supply security cooperation mechanism assuming voluntary participation and financial compensation. For this we formulate a stylized model, which allows the formal description and analysis of such a cooperation model. We demonstrate the operation of the proposed mechanism under the assumption of risk-averse participants and discuss possible critical issues related to implementation possibilities.

%According to the principles of the proposed framework, each participant reports (1) an inverse demand function, describing its price-flexibility; and (2) a value determining the level of cooperation in the redistribution process. After the resolution of uncertainty regarding the levels of accumulated gas in reservoirs and the availability of transfer capacities, gas is redistributed among participants in order to supply those, who need it most according to the reported inverse demand function. Participants from whom gas is relocated are compensated according to a redistribution-clearing price. The redistribution process takes into account the actual transmission constraints of the network and the levels of cooperation of participants. The level of participation limits the redistribution quantity of the participant in both directions: No participant may receive more gas than this quantity, and the quantity which is reallocated at the expense of the participant, is also limited by this quantity.

\section{The proposed solidarity mechanism}
\label{section_SM}
In this section we present the underlying key assumptions of the proposed mechanism, introduce the formal game and the corresponding network model. 
\subsection{Basic assumptions}
We study the interaction of strategic decision makers of countries, who aim to cover the energy demand of the country's economy. These entities, who will correspond to the players of the implied game may be best interpreted as national energy companies, who have access to storage facilities.

The proposed solidarity framework is interpreted in an environment, where every potential participant (player) individually bargains with external suppliers in order to fill its storage capacity for the winter period. However, the success of this bargaining is (at least partially) uncertain at the time of solidarity contracting. 

According to the proposed framework, the first phase of the mechanism, the solidarity contracting takes place in the spring period, before any player would begin to gather resources individually. In this period, every player has to decide whether they participate in the proposed cooperation framework, or not. The voluntary aspect of the proposed mechanism implies that each participant has the exclusive right to determine this value ($q^P$).
If a player $n$ decides positively, it defines its nonzero level of participation ($q^P(n)>0$). No participation corresponds to $q^P(n)=0$. The importance of this value is twofold. On the one hand it limits the quantity, which later may be taken away from the respective player to help others (in greater need), and at the same time it also limits the quantity which may be later received by the particular individual participant during the redistribution process.
If $q^P(n)>0$ applies, the participant also reports its expected demand in the form of piecewise constant inverse demand functions. Such functions
are able to differentiate between the various components of individual demand. It is assumed that these various components correspond to more (like residual heating) and less (industrial demand, community-related heating etc.) critical elements of the expected demand, characterizing the demand elasticity of the player in question \footnote{Note that revealing such information may negatively affect bargaining potential of participants in latter transactions. Thus, these reported date should be handled as confidential.}.

Phase two of the mechanism, the compensation, corresponds to the winter period, when higher demands and potential resource shortages arise.
We assume that at the beginning of the second phase, the following factors, which are still uncertain in phase one, are already determined and known for all participants: (1) The quantity of accumulated resource available to individual participants and (2) the actually available network transmission capacities. In the compensation phase, resources are redistributed among participants with non-zero participation levels according to individual needs, up to the determined levels of participation and considering the available transmission capacities. In other words, resources subject to the solidarity mechanism are routed to those participants who are in the highest need according to the demand data reported in phase one, taking into account network constraints. 
Players who have chosen not to the participate in phase one do not receive any additional resource during the redistribution process, but on the other hand, they will fully keep their accumulated resources.
Furthermore, if any non-zero redistribution transactions take place, the participants from whom resources are reallocated to other players are financially compensated, and the compensation price is determined according to the previously reported (inverse) demand functions.

The solidarity mechanism may be viewed as a special secondary market on which the participation is obligatory for those players who decided to define a nonzero level of participation. In the following we show how the elements of the above solidarity mechanism may be formalized using a computational model, and how such a model may provide insight into the potential operation and into the related strategic decisions in the case of such a solidarity framework.

\subsection{Formal game theoretic model}

It is easy to see that the potential benefit implied by the solidarity mechanism to any single player depends on the participation levels defined, and demand reported by other players as well (e.g. if only a single player participates, no redistribution may take place), thus the proposed supply-security related accumulation-redistribution process can be formally described as a game. For this we define the class of transaction-constrained resource-redistribution games under uncertainty (TCRRGU), and demonstrate the operation using an example. The proposed TCRRGU framework is based on strategic game in which, the strategic decision of the players correspond to the choice of $q^P(n)$ in phase one of the solidarity framework, while the payoffs of the players are determined in phase two, after the resolution of the uncertainties regarding resource accumulation and transmission capacities available for the redistribution.
We assume that every participating player provides its (future) natural gas demand in the form of a parametrized inverse demand function, based on which the latter redistribution processes takes place.

The redistribution process in phase two, which determines the outcome of the game, depends on the determined participation levels, on the accumulated resources and also on the available transport paths, which are subject to uncertainty at the time of solidarity contracting in phase one. For our model we assume that the nature and parameters of uncertainty are known to every player during the bargaining process (i.e. the uncertainty is structured, as described in subsection \ref{subsec_uncertainty}). In addition, as described earlier, a compensation process is defined related to the redistribution process.

\subsubsection{Network model}
The natural gas network is represented by a directed graph with $N$ nodes and $M$ edges where individual nodes represent players of the game.
Each edge $m$ is characterized by direction-dependent capacity values. $q^+ \in \mathcal{R}^M >0$ and $q^- \in \mathcal{R}^M <0$ denote the maximal transfer capacity vectors of edges in the positive and negative direction respectively. \footnote{Taking into account a sign convention, i.e. the maximal transferable quantity of edge $m$ in the negative direction is equal to $-(q^-(m))$. This allows us to use a single variable for the description of the flow on the edge, the sign of which defines the direction of the flow.}

The differentiation of transfer capacity over edge directions makes it possible to describe direction-dependent transfer capacity of pipelines, which may e.g. depend on the presence of compressor stations along the pipeline. 

\subsubsection{Consumer demand}
\label{consumer_demand}
We assume that players describe their elasticity of demand via piecewise constant inverse-demand functions composed of $W$ steps. Each piecewise constant part is characterized by two parameters: A price ($p^c$) and a consumption quantity ($q^c$). In this formalism $p^c_{n,j}$ denotes the price (per unit) level of the $j$-th step of the inverse demand function of player $n$, while $q^c_{n,j}$ defines the quantity (width) of the j-th step ($j \in \{1,...,W\}$).

\subsubsection{Uncertainty}
\label{subsec_uncertainty}
We take uncertainty into account on two levels. First, we assume that it is uncertain, how successful the players are in accumulating resources (i.e. filling storages to fully cover their later future demand) between phase one and phase two. Second, we assume that the state of the pipeline network in the context of available capacities for redistribution in phase two is also subject to uncertainty in the sense that technological factors (e.g. completion or delay of projects or possible faults) and external flows might limit the redistribution of resources -- we mean that these flows are 'external' in the context that they are not related to the supply-security cooperation.

Uncertainty is represented in the model by a finite number of '\emph{states of nature}' or '\emph{scenarios}', one of which is randomly realized at phase two. The total number of scenarios is equal to $S$. State of nature $s$ occurs with probability $p_s$ at the second phase and $\sum_{s=1}^{S} p_s=1$. This representation of uncertainty modelling is e.g. used in the financial literature discussing risk measures and allocations (see e.g. \cite{csoka2009stable}). In our model each scenario $s$ is characterized by an ordered tuple $(r_s,q^{-}_s,q^+_s)$, where $r_s \in \mathcal{R}^n$ defines the amount of resources available for the players in the case of the scenario $s$, while $q^- \leq q^-_s \leq 0$ and $0 \leq q^+_s \leq q^+$ define the available transfer capacities of edges in the case of the scenario $s$.

\subsubsection{The redistribution process}

Let us assume that the vector describing the levels of participation, denoted by $q^P\in \mathcal{R}^N$ is defined (the $n$-th element of the vector is $q^P_n$, the participation level of player $n$).
The redistribution process is described by an optimization problem, where the resources available in the particular scenario ($r_S$) which are subject to the solidarity mechanism are redistributed to maximize the total utility of gas consumption, according to the predefined inverse demand functions (i.e. gas is routed to those who need it most), while considering the participation levels and transmission constraints of the scenario. The linear program of the redistribution process in the case of scenario $s$ is described by the formulas (\ref{obj_fnc})-(\ref{constr_conservation}).

\begin{small}
\begin{align}
&\max_x gx ~~~\text{w.r.t.} \label{obj_fnc} \\
& -q^P(n) \leq \sum_{i\in E^{in}_n} f_i - \sum_{i\in E^{out}_n} f_i \leq q^P(n) ~~~~\forall ~n\label{constr_particip} \\
& \sum_{i\in E^{in}_n} f_i - \sum_{i\in E^{out}_n} f_i - \sum_w c_{n,w} + r_s(n)=0 ~~~~\forall ~n \label{constr_conservation}
\end{align}
\end{small}

The variable vector $x \in \mathcal{R}^{M+nW}$ of the problem is composed as

\begin{small}
\begin{equation}
x=\left(
    \begin{array}{c}
      f \\
      c \\
    \end{array}
  \right).
\end{equation}
\end{small}
$f \in \mathcal{R}^{M}$ denotes the vector of edge flows $f=[f_1,~...,f_M]^T$, where $f_m$ is the signed flow of edge $m$ ($q^-_s(m) \leq f_m \leq q^+_s(m)$), and
 $c \in \mathcal{R}^{nW}$ is the vector of consumptions $c=[c_{1,1},~c_{1,2},~...,c_{N,W}]$, where $c_{n,w}$ is the consumption related to the $w$-th step of the inverse demand function of player $n$ ($0\leq c_{n,w} \leq q^c_{n,w}$).

The objective function, described by (\ref{obj_fnc}), where $g$ is composed as  $g=[0^{1 \times M},~p^c_{1,1},~...,~p^c_{N,W}]$, is defined in order to maximize the consumption utility ($U^c$) of participants.
Constraint (\ref{constr_particip}) describes that the difference of the total inflow and outflow of any node is constrained by the respective participation level ($q^P$) from both above and below for each node $n$. In this formalism, $E^{in}_n$ and $E^{out}_n$ denote the set of incoming and outgoing edge indices of node $n$ respectively.
The conservation constraint (\ref{constr_conservation}) describes that the sum of inflows minus the sum of outflows must be equal to the consumption minus the available resources for each node $n$.

\subsubsection{Compensation}

During the redistribution process (if the outcome is non-trivial, i.e. a nonzero redistribution takes place, thus flows are arising in the network), the reserves (thus the consumption utility) of some participants are decreased, while those to whom the gas is redistributed gain additional consumption utility.
The resulting consumption utility ($U^c(n)$) of each player $n$ may be easily derived based on the respective inverse demand function, and the consumption values ($c_{n,w}$), as described by eq. (\ref{eq_Uc}).

\begin{equation}\label{eq_Uc}
  U^c(n)=\sum_w c_{n,w} p^c_{n,w}
\end{equation}

In the proposed framework it is assumed that the players receiving additional gas during the redistribution have to financially compensate those, who suffer a decrease in their reservoir levels. Accordingly, the financial utility of player $n$ $U^f(n)$ is defined simply as the amount of money they receive or pay in the redistribution process. This value is nonzero if and only if resource is allocated to them from other participants or rerouted from them to other players.
Furthermore, $\sum_n U^f(n)=0$.
The compensation is based on the redistribution clearing price ($RCP_s$ denotes the redistribution clearing price in the case of scenario $s$). $RCP_s$ is calculated as follows. If a nonzero redistribution takes place, the resulting consumption for at least a subset of players is not equal to the resources originally available for them. The marginal increment in $U^c(n)$ in the case of the scenario $s$ ($\mu_s(n)$) is defined as the consumption utility increment (or decrement) implied by the last unit of gas redistributed to (or from) $n$.

\subsubsection{Risk measurement and aversion}
\label{subsec_risk_measure}

Let us emphasize that the model elements described up to this point are the principles, which may serve as basis for a voluntary redistribution mechanism (based on the choice of $q^P(n)$ for each player $n$). In the following, we will be interested in considerations, based on which players may determine their level of participation ($q^P(n)$), and show that assuming risk-averse players, the framework indeed motivates players for participation.

In order to do this, we have to define how players of the game measure the risk in this context. For this aim, we will use the concept of expected shortfall (ES) \cite{acerbi2002coherence,adam2008spectral}, which is a coherent measure of risk \cite{artzner1999coherent}.
The $\alpha$-expected shortfall is calculated as the expected value of the worst $\alpha~\%$ of the scenarios.
In the context of the proposed game-theoretic model, the strategic aim of the players is to minimize the ES of their consumption utility values ($ES(U^C)$).

\section{Example}
\label{section_example}

\begin{figure}[h!]
  \centering
  \includegraphics[width=5cm]{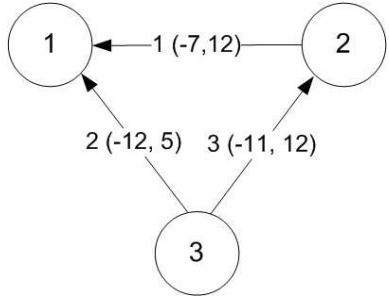}\\
  \caption{Simple example network network. Edges are labelled with their index $m$, and $q^-(m)$ and $q^+(m)$ in parentheses.}\label{Network_1}
\end{figure}

Figure \ref{Network_1} depicts the network of the considered simple example, with $N=M=3$, where the edge labels include the index of the edge ($m$), and the $(q^-(m),q^+(m))$ values. Nodes are labelled with their indices ($n$).

\begin{figure}[h!]
  \centering
  \includegraphics[width=5cm]{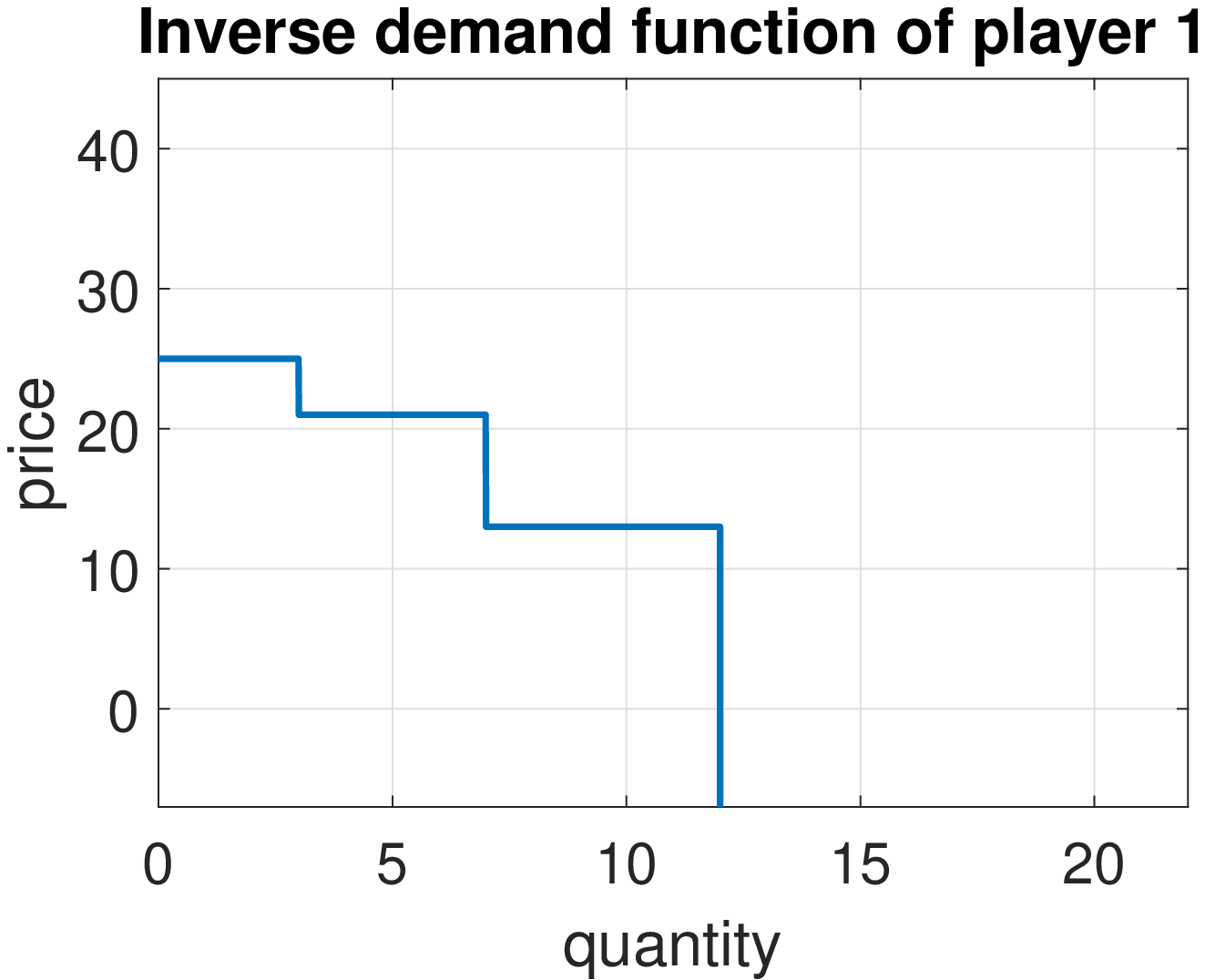}~
  \includegraphics[width=5cm]{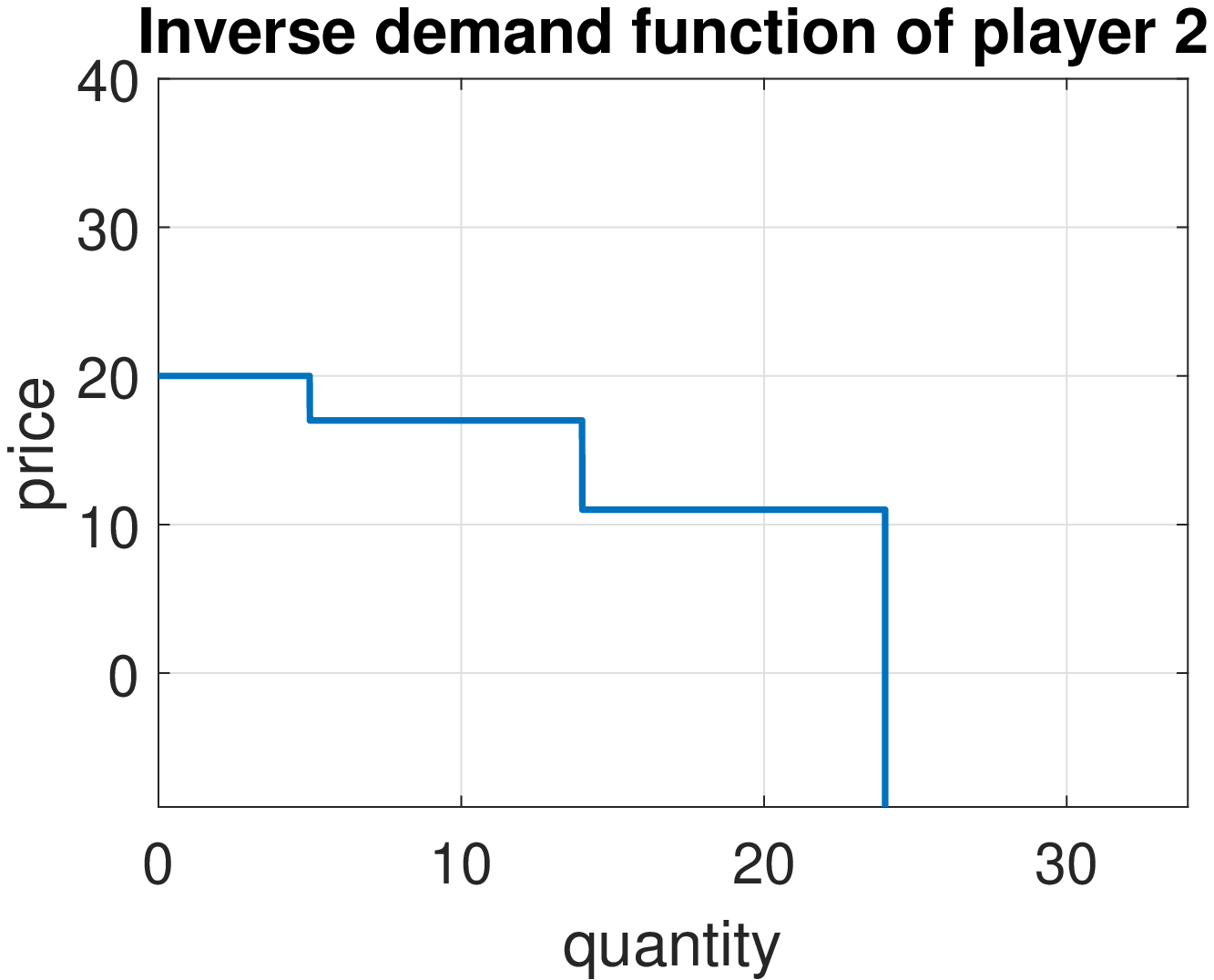}
  \includegraphics[width=5cm]{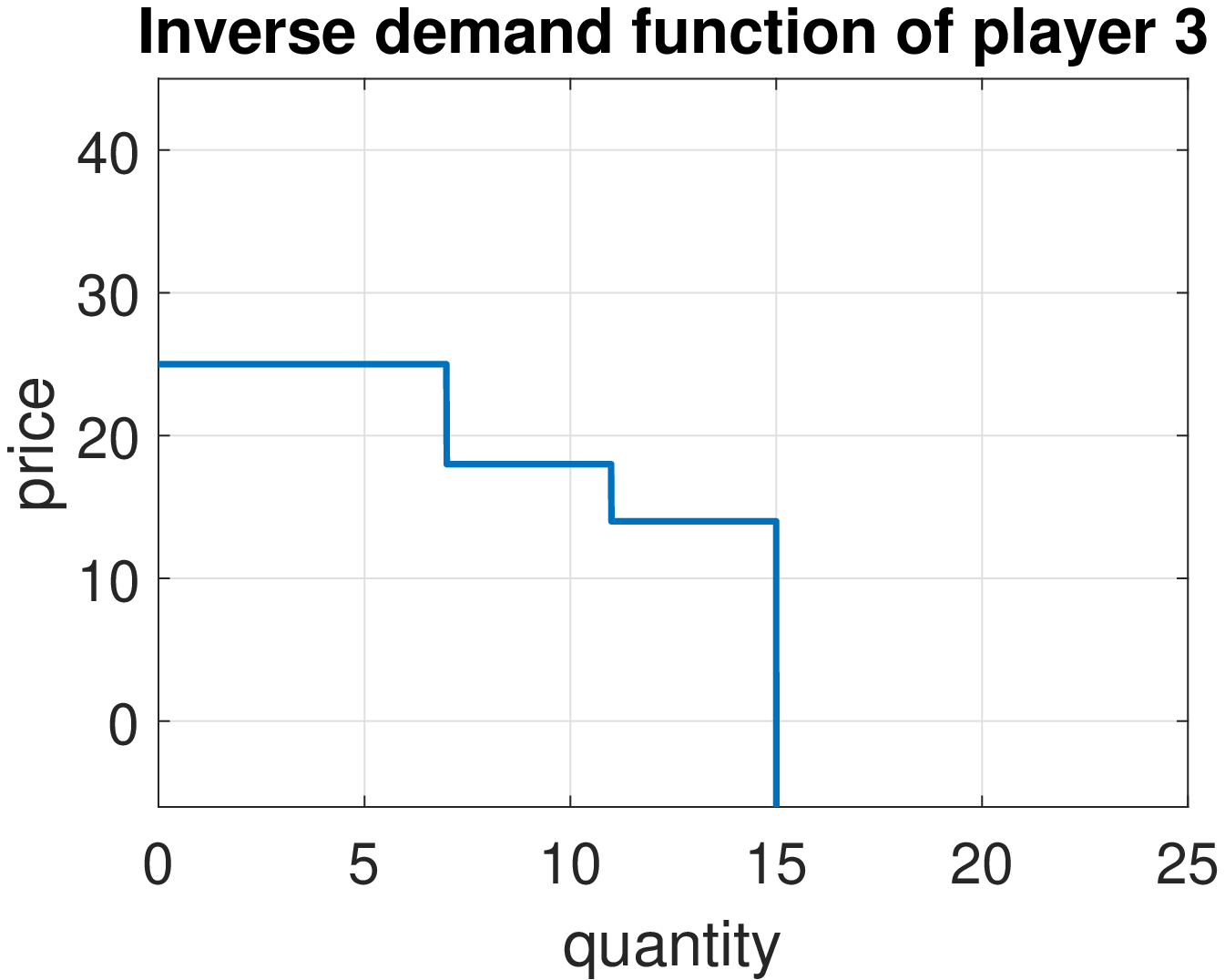}
  \caption{Inverse demand functions of consumers.}\label{Fig_IDF_1}
\end{figure}

The parameters of the demand functions used in the example and depicted ion Fig. \ref{Fig_IDF_1} are summarized in Table \ref{Table_IDF}.

\begin{table}[h!]
\begin{small}
\begin{center}
\begin{tabular}{|c|c|c|c|c|c||c|c|c|c|c|c|}
  \hline
  $p^c_{1,1}$ & 25 & $p^c_{2,1}$ & 20 & $p^c_{3,1}$ & 25 & $q^c_{1,1}$ & 3 & $q^c_{2,1}$ & 5 & $q^c_{3,1}$ & 7 \\ \hline
  $p^c_{1,2}$ & 21 & $p^c_{2,2}$ & 17 & $p^c_{3,2}$ & 18 & $q^c_{1,2}$ & 4 & $q^c_{2,2}$ & 9 & $q^c_{3,2}$ & 4 \\ \hline
  $p^c_{1,3}$ & 13 & $p^c_{2,3}$ & 11 & $p^c_{3,3}$ & 14 & $q^c_{1,3}$ & 5 & $q^c_{2,3}$ & 10 & $q^c_{3,3}$ & 4 \\
  \hline
\end{tabular}
\caption{Parameters of the inverse demand functions considered in the example.} \label{Table_IDF}
\end{center}
\end{small}
\end{table}

Regarding the uncertainty in the case of our simple example, let us assume $S=4$ and $p_s=0.25 ~~\forall s \in \{1 ,...,S\}$. Furthermore let the scenario parameters be as described in equations (\ref{eq_q_scen}) and (\ref{eq_r_scen}), where $q^+_s$ and $q^-_s$ are describing the available edge capacities in the case of scenario $s$ regarding positive and negative directions respectively, and $r_s$ is the vector of available resources (gas amount in the storage) in the case of the scenario $s$ ($r_s(n)$ corresponds to the resource of player $n$).

\begin{small}
\begin{align}
&q^-_1=\left(
        \begin{array}{c}
          -7 \\
          -3 \\
          -4 \\
        \end{array}
      \right)
~~
q^+_1=\left(
        \begin{array}{c}
          7 \\
          5 \\
          12 \\
        \end{array}
      \right)
~~
q^-_2=\left(
        \begin{array}{c}
          -5 \\
          -12 \\
          -7 \\
        \end{array}
      \right)
~~
q^+_2=\left(
        \begin{array}{c}
          12 \\
          0 \\
          12 \\
        \end{array}
      \right)
\nonumber \\
&q^-_3=\left(
        \begin{array}{c}
          -1 \\
          -12 \\
          -1 \\
        \end{array}
      \right)
~~
q^+_3=\left(
        \begin{array}{c}
          12 \\
          1 \\
          12 \\
        \end{array}
      \right)
~~
q^-_4=\left(
        \begin{array}{c}
          -5 \\
          -12 \\
          -6 \\
        \end{array}
      \right)
~~
q^+_4=\left(
        \begin{array}{c}
          12 \\
          0 \\
          12 \\
        \end{array}
      \right)
      \label{eq_q_scen}
      \\
 &     r_1=[ 12 ~~ 12 ~~ 12]~~~r_2=[ 6  ~~  21  ~~  13]~~~r_3=[  10 ~~   16  ~~   8]~~~r_4=[6  ~~  17 ~~   13]       \label{eq_r_scen}
\end{align}
\end{small}

%==

To give an example for the calculations of the compensation mechanism, assuming the inverse demand functions depicted in Fig. \ref{Fig_IDF_1} and summarized in Table \ref{Table_IDF}, if player 3 had originally 5 units of gas in the case of scenario $s$ ($r_s(3)=5$) and during the redistribution it receives an additional 5 units ($\sum_w c_{3,w}=10$), then $\mu_s(3)=18$, since the last unit of received gas implied increased $U^c(n)$ by 18 units. According to this, $RCP_s$ is defined as described in eq. (\ref{eq_RCP_s}).

\begin{equation}\label{eq_RCP_s}
  RCP_s= \frac{\min_{n: \mu_s(n)>0} \mu_s(n) + \max_{n: \mu_s(n)<0} (-\mu_s(n)) }{2}
\end{equation}

Thus, the financial utilities of players, denoted by $U^f$ are determined by transactions related to redistribution, cleared on the price of $RCP_s$.
As $RCP_s$ is always higher than the marginal consumption utility of any player from whom gas is rerouted to others, and always less than the marginal utility of any player receiving gas, the sum of the change in the consumption utility due to redistribution ($\Delta U^c$) and the financial utility ($U^f$) is always nonnegative for each player by construction.

\subsubsection{Example scenario calculation}

Before we discuss the questions related to the determination of $q^P$ values, to give an example, let us calculate the outcome of scenario 1, assuming $q^P=[2~~2~~4]$. According to eq. (\ref{eq_r_scen}), the available resources are equal to 12 units for each of the players in the case of this scenario. If we solve the optimization problem (\ref{obj_fnc})-(\ref{constr_conservation}) taking into consideration the inverse demand functions defined in table \ref{Table_IDF} (and depicted in Fig. \ref{Fig_IDF_1}), and the pipeline capacity constraints defined in eq. (\ref{eq_q_scen}), we obtain the solution, in which 2 units of gas from player 1 are redistributed to player 2, via edge 1 (implying a flow of -2 because of the opposite flow compared to the reference direction of the edge).
This implies that the vector of consumption utilities ($U^c$) changes from $[224~~219~~261]$ (the reference $U^c$ vector of the scenario, implied by $r_1$ without redistribution) to $[198~~253~~261]$ thanks to the redistribution process.
In this case $RCP_1=15$, implying $U^f=[30~~-30~~0]$.

We may calculate the other scenarios similarly. In scenario 2, two units of gas are redistributed from player 2 to players 1 and 3 (1 unit for each), implying a resulting $U^c=[159~~308~~289]$ and $U^f=[-12.5~~25~~-12.5]$. Regarding scenario 3, players 1 and 2 both provide 2 units of gas for player 3, resulting in
$U^c=[172~~253~~261]$ and $U^f=[27~~27~~-54]$. Finally, in scenario 4, two units of gas are redistributed from player 2 to players 1 and 3 (1 unit for each), implying a resulting $U^c=[159~~264~~289]$ and $U^f=[-12.5~~25~~-12.5]$.
Table \ref{Table_UcUf} summarizes the utility results for the various scenarios.

\begin{table}[h!]
\begin{small}
\begin{center}
\begin{tabular}{|c||c|c|c|c|}
  \hline
  scenario  & 1 & 2 & 3 & 4 \\ \hline
  $U^c(1)$ & 198 (224)  & 159 (138)& 172 (198) & 159 (138) \\
  $U^c(2)$ & 253 (219)  & 308 (330)& 253 (275) & 264 (286) \\
  $U^c(3)$ & 261 (261)  & 289 (275)& 261 (193) & 289 (275) \\
  $U^f(1)$ & 30        & -12.5    & 27        & -12.5 \\
  $U^f(2)$ & -30       & 25       & 27        & 25 \\
  $U^f(3)$ & 0         & -12.5    & -54       & -12.5 \\
  $RCP$    & 15        & 12.5     &  13.5     &  12.5 \\
  \hline
\end{tabular}
\caption{Resulting consumption and financial utilities and $RCP$ values in the example. In the rows of $U^c$ the original $U^c$ values without redistribution is indicated in parentheses.} \label{Table_UcUf}
\end{center}
\end{small}
\end{table}

Regarding to the risk measure defined in subsection \ref{subsec_risk_measure}, we assume that the players consider the value of $U^c$ as basis of their calculations, and in this simple example we use the $\alpha=25\%$ expected shortfall.
In this very case, the ES value of each player may be calculated as taking the worst $U^c$ value over the possible (4) scenarios.

We may compare the ES values of players, without or with cooperation, assuming $q^P=[0~~0~~0]$ and $q^P=[2~~2~~4]$, respectively.
Based on table \ref{Table_UcUf} we may easily conclude, that the ES values are increased with the cooperation: From 138 to 159, from 219 to 253 and from 193 to 261 in the case of player 1, 2 and 3, respectively. This increase in the ES values of players clearly shows that cooperation (with $q^P=[2~~2~~4]$ in this case) reduces the risk for the players.

\subsubsection{The implied strategic game}

Are the values $q^P=[2~~2~~4]$ the most efficient choices for players to reduce risk, i.e. to maximize their ES values in the current context? In general, the resulting ES value of player $n$ in the case of choosing $q^P(n)$ clearly depends on the choice of $q^P$ of other players (e.g. if other players choose $q^P=0$, it is sure that the cooperation will not bring any benefit for player $i$). Thus, the framework defines a nonzero-sum matrix game, where the strategy space of the players is given by their possible choices of $q^P$, and their payoff is their respective ES value in the resulting multi-player strategy space. If we constrain the possible set of choices for $q^P$ for integer values for simplicity, we can also state that $q^P=[2~~2~~4]$ is a Nash-equilibrium of this non-cooperative game.

\section{Discussion}
\label{sec_discussion}

\subsection{Equilibrium aspects}

The reader may ask, how the equilibrium point of the example has been determined, or in general, how is it possible to determine equilibrium point(s) of the implied non-cooperative game. Although the main aims of this article do not include to deeply discuss the equilibrium properties of the non-cooperative game class implied by the TCRRGU problem, let us formulate some observations.

Since the expected benefit of participation in the mechanism strongly depends on the defined participation quantities of other players, an iterative scheme (repeated game) may be used for the definition of the $q^P_n$ values, which potentially allows the players to reach a Nash equilibrium.

The iterative application of best-response strategies (see e.g. \cite{csercsik2015traffic}) potentially leads to equilibrium. In the case of the proposed example, the initial $q^P$ vector was determined based on the parameters of the inverse-demand functions, as $q^P(0)=[5~~10~~4]$, i.e. all players defined their initial $q^P$ value as the quantity parameter of the last step of their inverse demand function. Iterating the best response functions in this case led to an equilibrium after 3 steps.

Even in the case of the proposed simple example, this equilibrium is not unique. All of the vectors $q^P=[1~~1~~1]$, $q^P=[1~~2~~1]$ $q^P=[2~~2~~4]$ $q^P=[1~~2~~3]$ show the equilibrium property, i.e. no player $n$ is able to increase its own $ES$ value by unilaterally changing $q^P(n)$ (if we restrain ourselves on integer values regarding the level of participation).

\subsection{On incentive compatibility}

Let us return to a key assumption of the proposed framework for some discussion. The cooperation framework is based on the reported inverse demand functions, which are used in the objective function of the optimization steps. It is easy to see that if a player submits an inverse demand function with high price-parameters, it is likely that the gas will be redistributed for him/her in most of the scenarios. However, as we have seen the $RCP$ is also determined by marginal utilities, and such a strategy will likely result in the increase of the $RCP$, thus it implies more loss for the player through $U^f$ (since the gas is redistributed to him/her, he/she will have to pay for it). Let us emphasize that this does not mean at all that the proposed framework necessary motivates players to reveal the parameters of their true inverse-demand functions -- this, and more details related to incentive-compatibility \cite{nisan2007} may be the subject of later studies.

%\subsection{Full or partial cooperation structures}
%
%In the proposed example only three players have been considered.

\subsection{Potential practical implementation}

A further simplification of the proposed model is that it has been assumed that the possible scenarios and their realization probabilities are determined and known for each player. In practice, the uncertainty is much less exactly defined, various players potentially have different beliefs about it, and they calculate their risk measures and strategies according to these individual considerations. In other words, an abstract game such as the one proposed will never be realized in practice, which also implies that the theoretical properties of the proposed formal model shortly discussed above (equilibrium aspects and incentive compatibility) would have moderate significance in the case of a potential real-world application. Nevertheless, some elements of the proposed model and its solution concepts (like reporting of inverse demand functions, iterative determination of the levels of participation) may represent useful approaches in the process of designing realistic mechanisms in the future.

Altogether, our main aim in this article was to show that under textbook-like simplifying assumptions, such a cooperative using assuming a transparent mechanism and voluntary participation may indeed work, and such approaches may exhibit potential, when one would like to answer the question '\emph{How to improve supply-security related cooperation of the EU-member countries, and enable the more (internationally) efficient usage of storage facilities}'.

\section{Conclusions}
\label{sec_conclusions}

Regarding the usage of storage facilities, there are two possible ways to enhance the supply security of the EU for the following years.
(1) To constitute EU-level reserves and redistribution mechanisms, which aim to help the member states in potential future need. Since under the current circumstances, the construction/exploitation of new reservoirs and gas for this aim  do not seem to be not realistic in the short term, this approach would require the partial expropriation of national gas reserves and/or storage capacities. Such centralized approaches are likely to meet resistance by countries who consider that they made more sacrifices and effort for their own supply security in the past than others. We do not argue that such initiatives are necessary doomed, but it is possible that obtaining a sufficient political support for such a regulation framework will be challenging.
(2) The EU might also act as a catalyst of voluntary supply security cooperations by defining the appropriate transparent and predictable regulatory frameworks. Such approaches may complement or maybe even partially substitute the initiatives of the first type to further enhance the dynamism and flexibility of the reaction of the Union in the case of an emergency event. Based on simple computational modelling studies, this paper argues that multilateral voluntary supply security cooperation mechanisms may have significant potential in reducing the individual risk of participants.

%===============

\bibliography{RD.bib}% common bib file

\begin{thebibliography}{10}

\bibitem{acerbi2002coherence}
Carlo Acerbi and Dirk Tasche.
\newblock On the coherence of expected shortfall.
\newblock {\em Journal of Banking \& Finance}, 26(7):1487--1503, 2002.

\bibitem{adam2008spectral}
Alexandre Adam, Mohamed Houkari, and Jean-Paul Laurent.
\newblock Spectral risk measures and portfolio selection.
\newblock {\em Journal of Banking \& Finance}, 32(9):1870--1882, 2008.

\bibitem{araujo2018game}
Felipe~Costa Araujo and Alexandre~Bevilacqua Leoneti.
\newblock Game theory and 2x2 strategic games applied for modeling oil and gas
  industry decision-making problems.
\newblock {\em Pesquisa Operacional}, 38:479--497, 2018.

\bibitem{artzner1999coherent}
Philippe Artzner, Freddy Delbaen, Jean-Marc Eber, and David Heath.
\newblock Coherent measures of risk.
\newblock {\em Mathematical Finance}, 9(3):203--228, 1999.

\bibitem{csercsik2015traffic}
D{\'a}vid Csercsik and Bal{\'a}zs Sziklai.
\newblock Traffic routing oligopoly.
\newblock {\em Central European Journal of Operations Research},
  23(4):743--762, 2015.

\bibitem{csoka2009stable}
P{\'e}ter Cs{\'o}ka, P~Jean-Jacques Herings, and L{\'a}szl{\'o}~{\'A}
  K{\'o}czy.
\newblock Stable allocations of risk.
\newblock {\em Games and Economic Behavior}, 67(1):266--276, 2009.

\bibitem{holland2009strategic}
Alan Holland.
\newblock Strategic interaction in ratcheted gas storage.
\newblock In {\em 2009 7th IEEE International Conference on Industrial
  Informatics}, pages 244--249. IEEE, 2009.

\bibitem{holland2013equilibrium}
Alan Holland and Christopher Walsh.
\newblock An equilibrium analysis of third-party access to natural gas storage.
\newblock {\em The Journal of Energy Markets}, 6(2):3, 2013.

\bibitem{hubert2015pipeline}
Franz Hubert and Onur Cobanli.
\newblock Pipeline power: A case study of strategic network investments.
\newblock {\em Review of Network Economics}, 14(2):75--110, 2015.

\bibitem{jafarzadeh2021possibility}
Amir Jafarzadeh, Abbas Shakeri, Abdolrasoul Ghasemi, and Afshin Javan.
\newblock Possibility of potential coalitions in gas exports from the southern
  corridor to {E}urope: a cooperative game theory framework.
\newblock {\em OPEC Energy Review}, 45(2):217--239, 2021.

\bibitem{janjua2022asymmetric}
Shahmir Janjua, Muhammad~Umair Ali, Karam~Dad Kallu, Amad Zafar, Shaik~Javeed
  Hussain, Hasnain Gardezi, and Seung~Won Lee.
\newblock An asymmetric bargaining model for natural-gas distribution.
\newblock {\em Applied Sciences}, 12(11):5677, 2022.

\bibitem{kiely2016market}
Greg Kiely.
\newblock {\em A market consistent gas storage modelling framework: valuation,
  calibration, \& model risk}.
\newblock PhD thesis, University of Limercik, 2016.

\bibitem{nisan2007}
Noam Nisan, Tim Roughgarden, \'{E}va Tardos, and Vijay~V. Vazirani.
\newblock {\em Algorithmic Game Theory}.
\newblock Cambridge University Press, 2007.

\bibitem{rey2020regional}
Nikidrea~Toreece Rey.
\newblock {\em Regional competition and cooperation: game theory analysis of
  the "Convention on the Legal Status of the Caspian Sea"}.
\newblock PhD thesis, The University of Texas at Austin, 2020.

\bibitem{roson2015bargaining}
Roberto Roson and Franz Hubert.
\newblock Bargaining power and value sharing in distribution networks: A
  cooperative game theory approach.
\newblock {\em Networks and Spatial Economics}, 15(1):71--87, 2015.

\bibitem{schaef2014co2}
H~Todd Schaef, Casie~L Davidson, A~Toni Owen, Quin~RS Miller, John~S Loring,
  Christopher~J Thompson, Diana~H Bacon, Vanda~A Glezakou, and B~Pete McGrail.
\newblock Co2 utilization and storage in shale gas reservoirs: experimental
  results and economic impacts.
\newblock {\em Energy Procedia}, 63:7844--7851, 2014.

\bibitem{schitka2014applying}
Barrett~B Schitka.
\newblock Applying game theory to oil and gas unitization agreements: how to
  resolve mutually beneficial, yet competitive situations.
\newblock {\em The Journal of World Energy Law \& Business}, 7(6):572--581,
  2014.

\bibitem{Toufighi2022}
Seyed~Pendar Toufighi.
\newblock Assessing the stability of the oil and gas production in common
  fields: Application of game theory.
\newblock {\em Journal of Economics, Finance and Management studies},
  5:1250--1262, 05 2022.

\bibitem{ENTSOG2022}
Kacper Zeromski, Louis Watine, and Jacques Reberol.
\newblock {ENTSOG Yearly Supply Outlook 2022}.
\newblock
  \url{https://www.entsog.eu/sites/default/files/2022-07/SO0036-22_Yearly_Supply_Outlook_2022-2023_0.pdf},
  2022.
\newblock {European Network for Transmission Operators for Gas}.

\bibitem{zhang2007system}
Yingqi Zhang, Curtis~M Oldenburg, Stefan Finsterle, and Gudmundur~S Bodvarsson.
\newblock System-level modeling for economic evaluation of geological co2
  storage in gas reservoirs.
\newblock {\em Energy Conversion and Management}, 48(6):1827--1833, 2007.

\end{thebibliography}

\end{document}